# The Effect of Temperature on Amdahl Law in 3D Multicore Era

L. Yavits, A. Morad, R. Ginosar

**Abstract**—This work studies the influence of temperature on performance and scalability of 3D Chip Multiprocessors (CMP) from Amdahl's law perspective. We find that 3D CMP may reach its thermal limit before reaching its maximum power. We show that a high level of parallelism may lead to high peak temperatures even in small scale 3D CMPs, thus limiting 3D CMP scalability and calling for different, in-memory computing architectures.

———————————— ◆ ————————————

## 1 INTRODUCTION AND RELATED WORK

Power consumption is among the main factors limiting the scalability of Chip Multiprocessors (CMP) [19]. As integration driven by device scaling slows down [11], three-dimensional (3D) integration arises as a natural step in CMP evolution. 3D CMP implementation has the potential to speed up the sequential portion of the code, lifting the main limiting factor identified by Amdahl's law [20]. 3D integration mitigates the off-chip memory bandwidth restrictions and enhances connectivity by stacking one or multiple DRAM layers above CMP layers, and enabling TSV based vertical communication. A conceptual 3D CMP, with cores partitioned into a number silicon layers, featuring an embedded multilayer 3D DRAM is presented in Fig. 1.

Unfortunately, 3D integration cannot eliminate the 'power wall'. With power scaling slowing down, stacking a number of CMP core layers necessarily results in a significant increase of power density. Growing power density leads to higher temperatures, which affect the performance and reliability of 3D designs. For example, placing DRAM above CMP layers might be thermally prohibitive because of hot spots where temperature rises above the DRAM operational range (up to 95°C), such as in 3D DRAM cache [3].

A classical CMP architecture paradigm includes design choices such as symmetric *vs.* asymmetric CMP [18], number of cores *vs.* core size [18], cores *vs.* cache [1][14] etc. When designing a 3D CMP, the computer architect must address an additional question: How does the temperature affect the number of cores of 3D CMP and their size? This paper strives to answer this question and quantify the impact of the thermal aspects of 3D design on the performance and scalability of CMP.

In recent years, there has been an extensive research into corollaries of Amdahl's law in the era of CMP. Hill and Marty [18] introduced an upper-bound analytical model for the performance and scalability of multicore and suggested an extension of Amdahl's law. Woo and Lee [4] extended the multicore performance and scalability model by addressing power consumption. Cassidy and Andreou [1] further developed the framework to account for optimal area allocation between core and memory, while Loh [8] extended Hill and Marty's model by adding the cost of the "uncore" resources. Chung *et al.* [5] extended the multicore corollary of Amdahl's law for heterogeneous architectures (including accelerators, such as FPGA, ASIC or GPU in addition to conventional processing cores). Eyerman and Eeckhout [22] augmented Amdahl's law by including execution of critical sections. We studied the effects of communication and synchronization on performance and scalability of a multicore [15]. Wang and Skadron [13] added supply voltage and operating frequency to Hill and Marty performance model. Recently, Ananthanarayanan *et al.* [8] extended Amdahl's law to account for process variations.

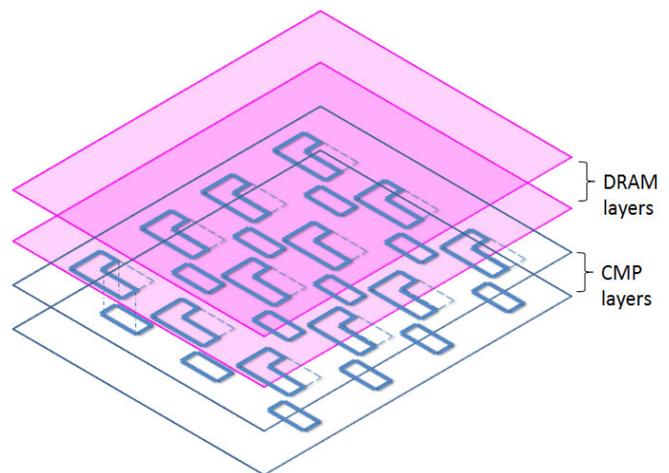

Fig. 1. 3D CMP with a 3D DRAM cube stacked above it.

In this work, we study the thermal effects of 3D integration on performance and scalability of a multicore from Amdahl's law perspective, a subject not addressed by prior research into corollaries of Amdahl's law in multicore era.

----

- *Leonid Yavits (\*), E-mail: yavits@tx.technion.ac.il.*
- *Amir Morad (\*), E-mail: amirm@tx.technion.ac.il.*
- *Ran Ginosar (\*), E-mail: ran@ee.technion.ac.il.*

*(\*) Authors are with the Department of Electrical Engineering, Technion-Israel Institute of Technology, Haifa 32000, Israel.*

We focus on qualitative trends rather than on actual temperature measurements. Using analytical modeling, we show that peak temperatures of 3D CMP grow with parallelism and number of cores. We further show that 3D CMP may reach a temperature limit before reaching the power constraint. As a result, a practical 3D CMP configuration is limited to a smaller number of larger cores. While actual temperature limit may change following innovations in thermal package development, this conclusion endures. We verify the results of analytical modeling by HotSpot [12] simulation using its default settings.

The rest of this paper is organized as follows. Section 2 presents the analytical model of 3D CMP temperature. Section 3 validates the analytical model by simulation. Section 4 offers conclusions.

## 2 ANALYTICAL MODEL OF 3D CMP TEMPERATURE

We consider a multicore with a constrained area budget enabling 256 "baseline core equivalent" (BCE [18]) cores, where core is a processing unit with its private cache(s). Following Hill and Marty, we consider symmetric and asymmetric CMPs, that is, all cores have the same ISA. We follow the power methodology of [4] and [5], as explained below. Let $p_{256}$ be the dynamic power consumption of a "full blown" processor of size $R = 256$ BCE (consuming the entire chip area). The power of a smaller core, of size $1 \leq r \leq 256$, is $p_r$. We further assume that runtime is long enough for the temperature to converge.

Ribando and Skadron [21] assume that power consumption of a large scale multicore, or manycore (two hundred cores) is identical to that of a small scale multicore (two cores) of the same area, and conclude that peak temperature of manycore is lower since its power distribution is more uniform. We take a different approach, based upon findings by Grochovsky and Annavaram [6], that power consumption of computing core exhibits sublinear growth as its area increases.

The phenomenon of reduction in computing core's activity (power per area unit) as its size grows has a number of underlying reasons. Larger cores normally have larger cache (thus potentially improving the hit rate and the overall performance of the core), however RAM has lower transistor activity compared to logic. Larger cores are likely to be superscalar; however the multiple issue pipelines would not always be fully utilized, thus leading to lower transistor activity. Larger cores may employ elaborate out-of-order execution structures, the efficiency of which depends on workload specifics. Finally, larger cores are likely to comprise hardware accelerators, which may improve the core performance but are not likely to be used continuously.

Following [5] and [6], we scale $p_r$ as a power law of the core size $r$:

$$p_r = p_{256} \cdot \left(\frac{r}{R}\right)^\alpha \quad (1)$$

where $\alpha = 0.875$ (according to [6], $Core\ Power = Core\ Performance^{1.7\ 5}$; substituting Pollack's rule ($Core\ Performance = Core\ Area^{0.5}$) [7] into this equation, we receive $Core\ Power = Core\ Area^{0.8\ 7\ 5}$).

The absolute temperature in a symmetric CMP (having $n_c$ cores of size $r$BCE each) can be written as follows [23]:

$$T = R_{TH}\frac{n_c p_r}{R} + T_a \quad (2)$$

where $R_{TH}$ is the thermal resistance (in units of Area·Kelvin/Watt), and $T_a$ is ambient temperature. We limit our analytical model to fully parallel programs (the parallelizable fraction of a program $f = 1$), where all cores are active. When $f < 1$, some of the cores are idle, which complicates the analytical model. Another reason for limiting the analytical model to $f = 1$ is that for any core size, the highest CMP temperature is reached when $f = 1$ (Section 3) and thus the $f = 1$ model provides a temperature upper bound.

The absolute temperature in an asymmetric CMP design with the serial core of size $r$ and $(R - r)$ 1BCE parallel cores [18] can be written as follows:

$$T = R_{TH}\frac{p_r + (R - r)p_1}{R} + T_a \quad (3)$$

The first component reflects the contribution of the serial core, while the second component reflects the parallel cores.

Fig. 2 and Fig. 3 present the absolute temperature of the symmetric and asymmetric CMP respectively, as a function of the core size $r$, for different values of $\alpha$.

The case of $\alpha = 1$ provides a lower bound on temperature (the entire chip consumes a fixed level of power regardless of core size, as in Fig. 2 and Fig. 3), and there is no room for any optimization. For $\alpha < 1$, the 3D CMP temperature drops with fewer larger cores and increases with larger number of smaller cores. The smaller the $\alpha$, the steeper the temperature change as a function of core size. These temperature changes happen more gradually for asymmetric CMP than for symmetric ones (as is evident when comparing Fig. 2 and Fig. 3).

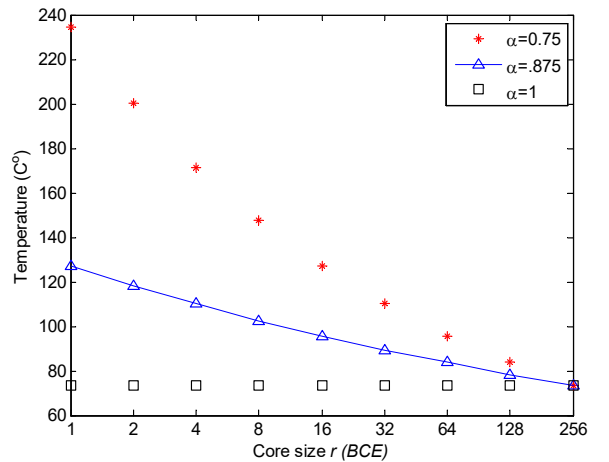

Fig. 2. Analytical model of symmetric CMP absolute temperature vs. core size $r$ for different $\alpha$, $R = 28mm^2$, $p_{256} = 25W$, $T_a = 20°C$ and $R_{TH} = 60mm^2K/W$.

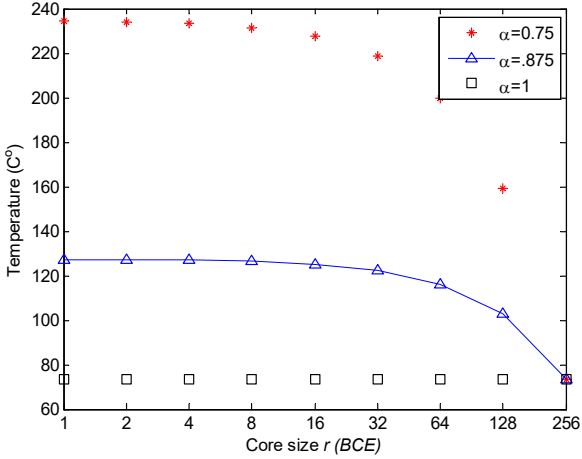

Fig. 3. Analytical model of asymmetric CMP absolute temperature vs. core size $r$ for different $\alpha$, $R = 28mm^2$, $p_{256} = 25W$, $T_a = 20°C$ and $R_{TH} = 60mm^2K/W$.

Assuming that the peak temperature of a CMP is limited (for example in 3D DRAM design, or simply due to package limitations), the thermal effect of the number of CMP cores and their size is an important design factor that must be addressed by computer architects. This aspect is discussed in the following section and Fig. 4.

## 3 VALIDATION OF THE ANALYTICAL MODEL

We validate our analytical model using HotSpot simulation [12] at its default configuration. The simulation parameters, heat sink assumptions, and description of the 3D stack are detailed at the end of this section.

We simulate a symmetric 3D CMP partitioned into 4 silicon layers with a DRAM layer integrated above them (the asymmetric case yields similar results). We use Intel's Nehalem 45nm processor as a reference full blown core with $p_{256} = 25W$ and $R = 28mm^2$ [11]. Hence, each layer of the simulated 3D CMP has a 256BCE budget of $R = 28mm^2$, divided into $n_c = R/r$ cores, each with an area of $r$BCE. No longer limited by the complexity of the model, we perform the simulation for different values of $f$.

Execution consists of two portions, serial and parallel. During serial execution, only one core of the multicore is active, while the other cores are idle. An idle core dissipates a fraction $k_r$ ($0 \leq k_r \leq 1$) of the power of an active core. Following [4], we assume $k_r = 0.2$.

During serial execution, the serial core dissipates $p_r$ power and $n_c - 1$ cores dissipate $p_r k_r$ power each. Execution time of this serial portion is $T_{serial} = T_r(1-f)/\sqrt{r}$, where $T_r$ is the sequential execution time on a single $r$BCE core, and $\sqrt{r}$ is a performance scaling factor based on Pollack's rule [7].

During parallel execution, all cores are active. Each core consumes $p_r$ power, and the execution time is $T_{parallel} = fT_r/(n_c\sqrt{r})$.

The inputs to HotSpot are the multicore floorplan and the power trace as described above.

Fig. 4(a) shows peak temperature of a symmetric 3D CMP as a function of its core size $r$ for different values of $f$. The figure also includes, as a reference, the maximum temperature allowed for DRAM. The peak temperature for $f = 1$ (measured in the center of the CMP layout) monotonically grows with the decreasing core size (increasing number of cores), generally in line with the analytical model results of Fig. 2. The peak temperature for $f < 1$ however first grows but then declines as the core size decreases. The peak CMP temperatures for $f < 1$ are consistently lower than for $f = 1$ (validating that $f = 1$ provides a temperature upper bound).

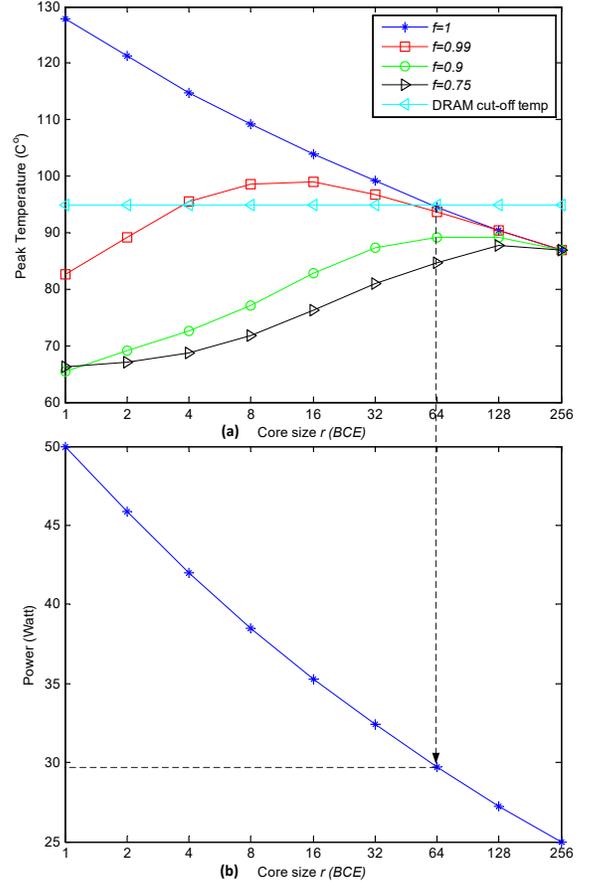

Fig. 4 (a) HotSpot simulated peak temperature vs. core size $r$ for different $f$ and $\alpha = 0.875$; (b) 3D CMP power consumption vs. core size, $f = 1$, $\alpha = 0.875$

Our findings are presented in the following three results.

Result 1: In highly parallelizable applications ($f = 1$), peak temperatures of 3D CMP grow with parallelism and the number of cores. 3D thermal constraints limit the CMP scalability, restricting the practical CMP configuration to a smaller number of larger cores. CMPs with a very large number of small active cores are less suitable for 3D implementation when $f = 1$.

For low level of parallelism (low thread count) the peak temperature of the CMP layer may be safely contained within the DRAM operational range. In our study, this happens for $f < 0.99$, but this figure may vary with the

CMP area and power parameters. However, for highly parallel tasks ($f \geq 0.99$ in our study), when most cores are active most of the time, the peak temperature of the CMP layers may reach beyond the DRAM operational range. A 3D implementation where DRAM cannot be placed atop a multi-layer CMP may fail one of its essential purposes, which is mitigating the bandwidth wall.

Result 2: Thermal considerations may constrain the scalability of 3D CMP due to thermal requirements of 3D DRAM integration.

Fig. 4(b) shows the 3D CMP power consumption as a function of its core size. As 3D CMP reaches its thermal limit for $f = 1$ (at 95℃ due to 3D DRAM integration), its power is short of 30W (as marked by the vertical dashed line connecting the two charts in Fig. 4(a), (b)), much lower than a typical Thermal Design Power (TDP) of contemporary high-end processors. Even for the thermal limit of a standard chip package, typically at 125℃, the 3D CMP power consumption for $f = 1$ is short of 50W, which is also below typical TDP.

Result 3: 3D CMP may reach its temperature limit (imposed for example by DRAM integration above the core layers, or package limitation) before exceeding its power limit.

Implication: Increasing parallelism (as suggested by Hill and Marty in [18]) in 3D CMP without addressing its thermal outcome has an adverse effect on 3D CMP scalability and performance. Hence, multicore designers should seek ways to reduce the activity of processing cores without decreasing their performance, to enable an efficient 3D integration of a large scale CMPs. Heterogeneous CMP [8] is one possibility. A radically different, non-CMP 3D architecture employing associative processing has also been shown to deliver high performance while maintaining temperatures compatible with 3D DRAM integration [16].

Note that when $f$ is lower than 0.99, Fig. 4(a) indicates that there are no thermal limitations on 3D integration with DRAM. However, asymmetric or heterogeneous architectures may be more appropriate in such cases [2].

Hotspot simulation parameters are presented in TABLE 1.

TABLE 1
HOTSPOT SIMULATION PARAMETERS

| Parameter | Value |
|---|---|
| Chip thickness | 0.15 mm |
| Convection capacitance | 140.4 J/K |
| Convection resistance | 0.1 K/W |
| Heat sink side | 60 mm |
| Heat sink thickness | 6.9 mm |
| Spreader side | 30 mm |
| Spreader thickness | 1 mm |
| Chip-to-spreader interface-material thickness | 0.02 mm |

Heat sink parameters are summarized in TABLE 2.

TABLE 2
HEATSINK PARAMETERS

| Parameter | Value |
|---|---|
| Convection | Forced |
| Flow type | Lateral airflow from sink side |
| Sink type | Fin-channel |
| Fin Height | 3 cm |
| Fin Width | 1 mm |
| Channel width | 2 mm |

The simulated 3D stack is based on [10] and depicted in Fig. 5.

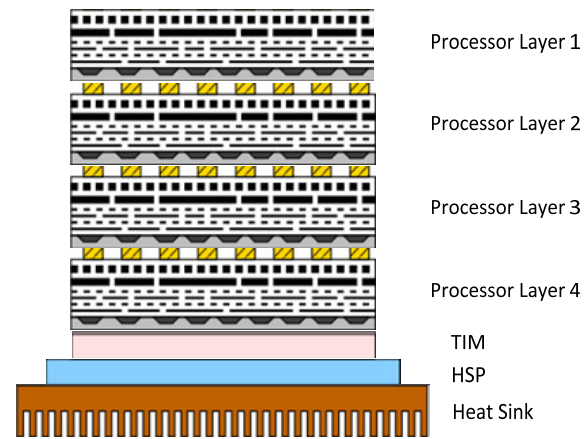

Fig. 5. Simulated 3D stack, based on [10].

## 4 CONCLUSIONS

As integration driven by device scaling slows down and the bandwidth wall looms, 3D integration becomes a natural step in CMP evolution. However, 3D designs are highly influenced by thermal aspects, not addressed by prior research into corollaries of Amdahl's law in multicore era. This work studies the effect of 3D CMP temperature on its performance and scalability from the perspective of Amdahl's law.

We find that the peak temperatures of 3D CMP grow with the number of cores and with task parallelism, potentially reaching a thermal limit before the power constraint of the CMP is reached. We also find that the peak temperature of 3D CMP may exceed the DRAM operational limit, thus making 3D DRAM integration difficult. Hence, the scalability of 3D CMP might be limited by thermal considerations, pushing the practical CMP configuration towards a smaller number of larger cores. CMPs with a large number of small active cores, targeted for highly parallelizable applications, may be less suitable for 3D integration.

An implication of our research is that increasing parallelism (as suggested by Hill and Marty in [18]) in 3D CMP without addressing its thermal outcome has an adverse effect on 3D CMP scalability and performance. Heterogeneous or non-CMP in-memory computing architectures [16] may prove more suitable for massively parallel 3D designs.

## ACKNOWLEDGMENT

This research was partially funded by the Intel Collaborative Research Institute for Computational Intelligence and by Hasso-Plattner-Institut.